\documentclass{article}
\usepackage{spconf,amsmath,graphicx}
\usepackage{enumitem}
\usepackage{booktabs}
\usepackage{xcolor,colortbl}
\usepackage{amsmath,amsthm}
\usepackage{multirow}
\usepackage{lipsum}
\usepackage{hyperref}
\newcommand{\myparagraph}[1]{\vspace{1.5ex}\noindent\textbf{#1{\hspace{1.5ex}}}}
\definecolor{Gray}{gray}{0.95}

\let\OLDthebibliography\thebibliography
\renewcommand\thebibliography[1]{
  \OLDthebibliography{#1}
  \setlength{\parskip}{0pt}
  \setlength{\itemsep}{5pt plus 0.3ex}
}

\title{Adapter-Based Extension of Multi-Speaker \\ Text-To-Speech Model for New Speakers}
\name{$^{*}$Cheng-Ping Hsieh$^1$\thanks{*Work done as an intern at NVIDIA.} \qquad Subhankar Ghosh$^2$ \qquad Boris Ginsburg$^2$}
\address{$^1$University of California, San Diego \\ $^2$NVIDIA, Santa Clara}

\begin{document}
%
\maketitle

\begin{abstract}
Fine-tuning is a popular method for adapting text-to-speech (TTS) models to new speakers. However this approach has some challenges. Usually fine-tuning requires several hours of high quality speech per speaker. There is also that fine-tuning will negatively affect the quality of speech synthesis for previously learnt speakers. In this paper we propose an alternative approach for TTS adaptation based on using parameter-efficient adapter modules. In the proposed approach, a few small adapter modules are added to the original network. The original weights are frozen, and only the adapters are fine-tuned on speech for new speaker. The parameter-efficient fine-tuning approach will produce a new model with high level of parameter sharing with original model. Our experiments on LibriTTS, HiFi-TTS and VCTK datasets validate the effectiveness of adapter-based method through objective and subjective metrics. 

\end{abstract}

\begin{keywords}
Text-To-Speech, Speaker Adaptation,  Adapter, Few-Shot Learning
\end{keywords}
\section{Introduction}
Neural text-to-speech (TTS) models have considerably improved in recent years with deep learning techniques \cite{wang2017tacotron, ping2017deep,ren2019fastspeech, ren2020fastspeech2, lancucki2021fastpitch} .
These models are capable of synthesizing natural human voice after being trained on several hours of high-quality single-speaker \cite{ljspeech17} or multi-speaker \cite{libritts, vctk, hifitts} recordings. However, to adapt new speaker voices, these TTS models are fine-tuned using a large amount of speech data, which makes scaling TTS models to a large number of speakers very expensive. 

\begin{figure}
    \centering
    \includegraphics[width=0.45\textwidth]{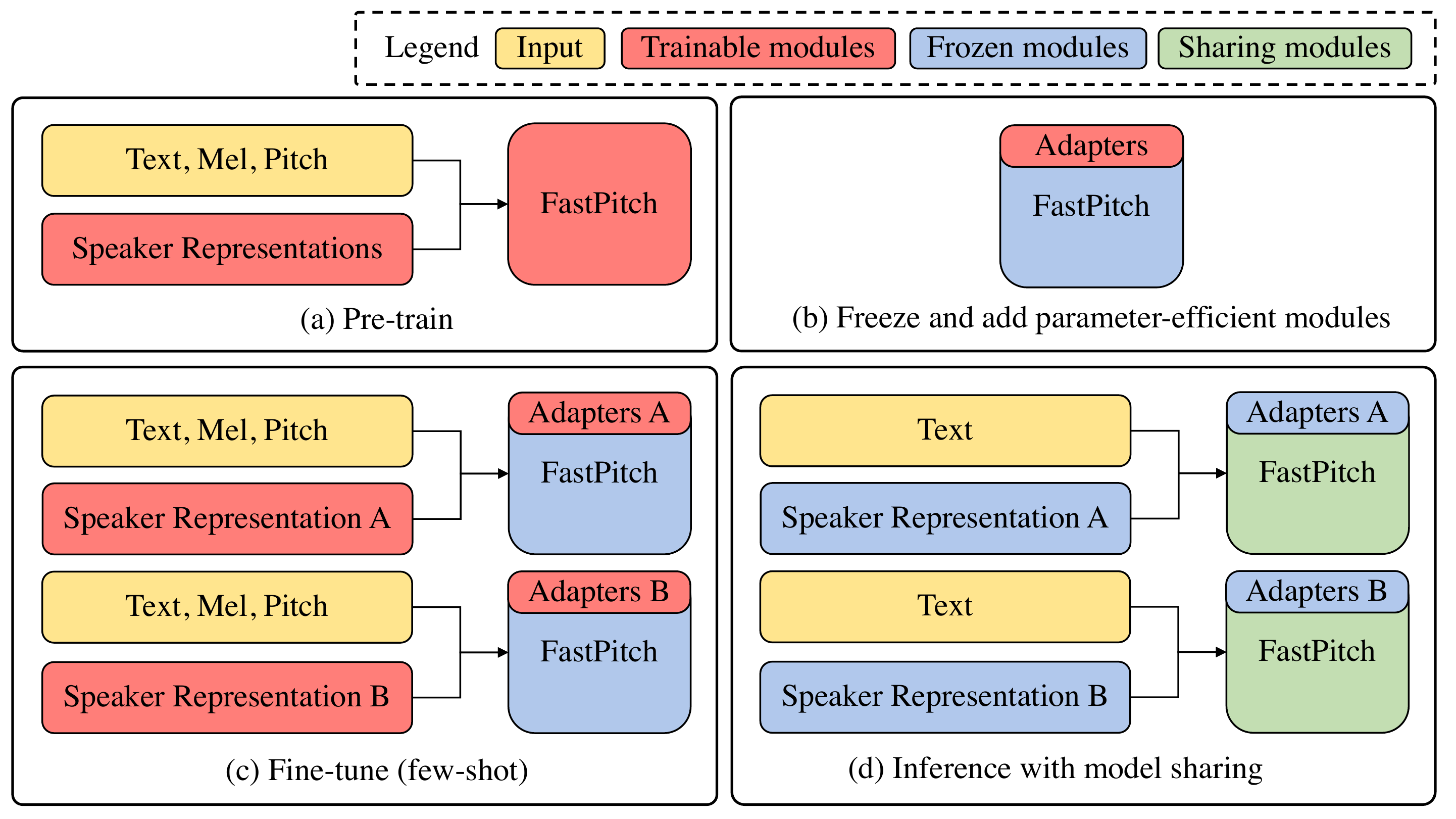}
    \vspace{-10pt}
    \caption{The proposed pipeline for adaptation of multi-speaker TTS model for new speakers. (a) Pre-train a multi-speaker FastPitch model. (b) Freeze weights of pre-trained FastPitch model and add adapter modules. (c) Only the adapters are fine-tuned for new speaker. (d) Inference by sharing the same model and plugging the lightweight, speaker-specific module.}
    \label{fig:pipeline}
    \vspace{-13pt}
\end{figure}

Fine-tuning of TTS models to new speakers may be challenging for number of reasons. First,the original TTS model should be pre-trained with a large multi-speaker corpus to make models to generalize well to new voices and different recording conditions.
Second, fine-tuning whole TTS model is very parameter inefficient, since one will need a new set of weights for every newly adapted speakers. 
Currently, there are two approaches to make adaptation of TTS more efficient. The first approach is to  modify only parameters directly related to speaker identity  \cite{moss2020boffin, zhang2020adadurian,  arik2018neural, chen2021adaspeech}. 
The other alternative approach is based on using a light voice conversion post-processing module to baseline TTS model \cite{voicefilter}.
The third challenge is to reduce amount of speech required to add  new speaker to existing TTS model.



In this paper, we propose a new parameter-efficient method for tuning existing multi-speaker TTS for new speakers shown in Fig.~\ref{fig:pipeline}. 
First, we pre-train a base multi-speaker TTS model on a large and diverse TTS dataset.
To extend model for new speakers, we add a few adapters -- small modules to the base model. We used vanilla adapter \cite{houlsby2019adapter}, unified adapters \cite{hu2021lora, li2021prefix, he2021unified}, or BitFit \cite{zaken2021bitfit}. Then, we freeze the pre-trained model and fine-tune only adapters on new speaker data. 
The contributions of this paper are:
\begin{itemize}[noitemsep, nolistsep, leftmargin=0.35cm]
\item 
We propose a new adapter-based  framework for efficient tuning of TTS model for new speakers without forgetting previously learned speakers.
\item  We validate our design through comprehensive ablation study across different types of   adapters modules, amounts of training data, and recording conditions.
\item We demonstrate that adapter-based TTS tuning performs similarly to full fine-tuning while demanding significantly less compute and data.
\end{itemize}
\vspace{-7pt}

\section{Method}
\begin{figure}
    \centering
    \includegraphics[width=0.45\textwidth]{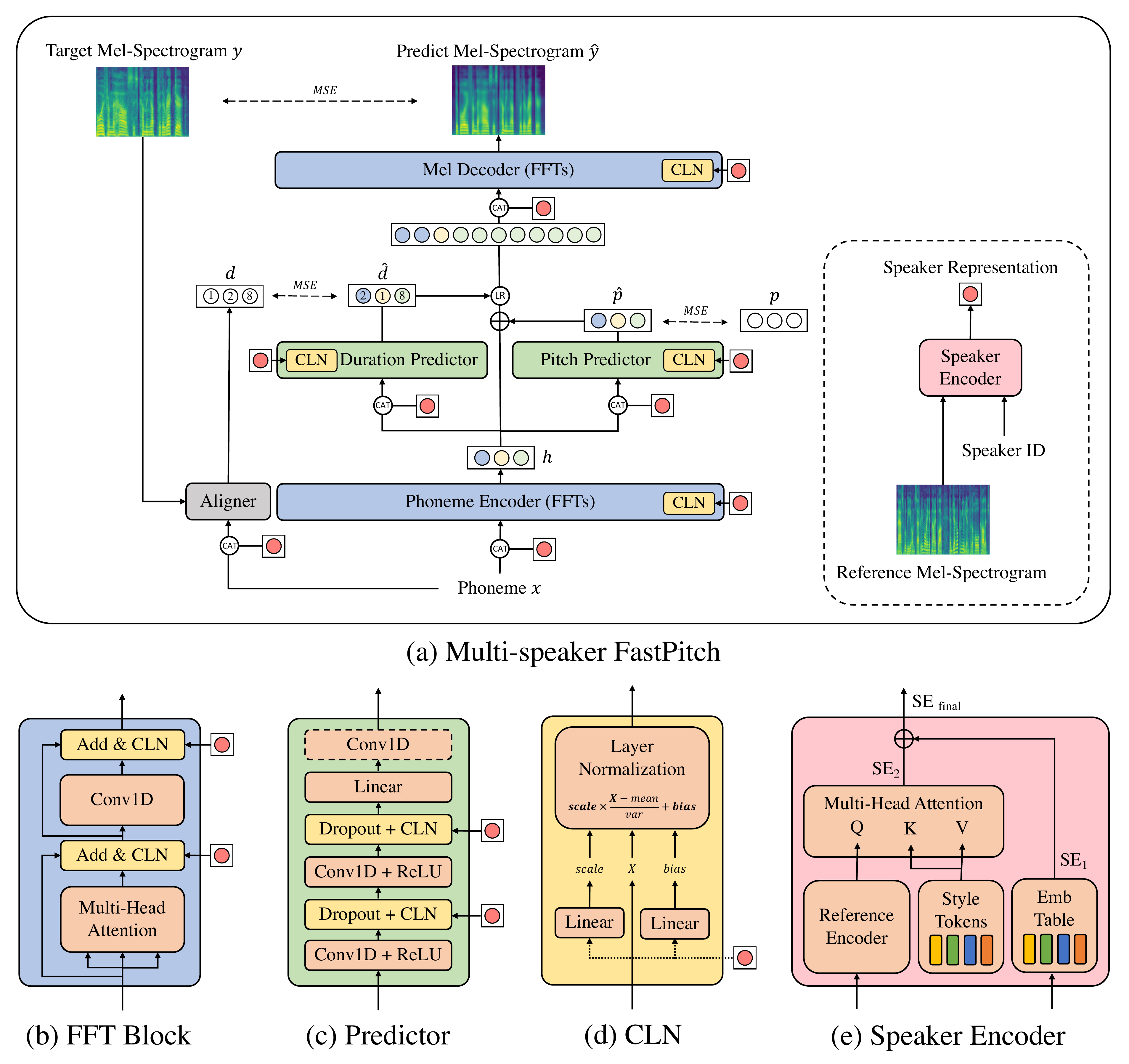}
    \vspace{-10pt}
    \caption{Architecture of proposed multi-speaker FastPitch. It is composed of phoneme encoder, mel decoder, duration and pitch predictor, aligner, and speaker encoder. We control speaker identity by using conditional layer normalization (CLN) and concatenating inputs with speaker representation.}
    \label{fig:fastpitch}
    \vspace{-15pt}
\end{figure}

\begin{figure}
    \centering
    \includegraphics[width=0.48\textwidth]{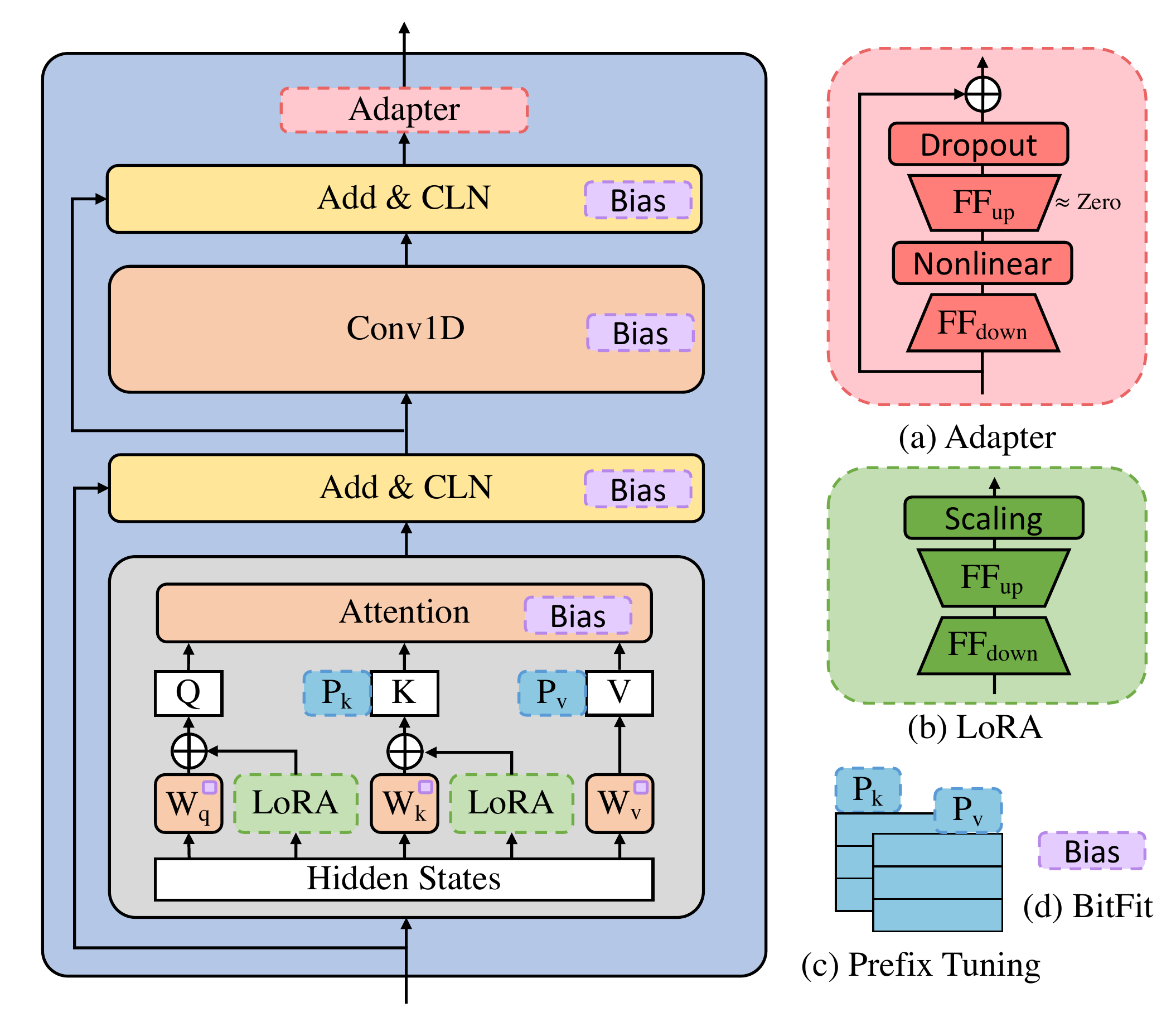}
    \vspace{-14pt}
    \caption{Illustration of parameter-efficient tuning modules in transformer architecture. LoRA and Prefix Tuning are only used in FFTs while Adapter and BitFit can be applied to any components in FastPitch.}
    \vspace{-10pt}
    \label{fig:adapter}
\end{figure}

In this section, we first describe the architecture of our pre-trained multi-speaker FastPitch --  a non-autoregressive TTS model conditioned on speaker representations, as shown in Fig.~\ref{fig:fastpitch}. Next, we introduce parameter-efficient adapter modules including vanilla adapter, unified adapters, and BitFit (see Fig.~\ref{fig:adapter}). Finally, we explain how we select the lightweight learnable modules to fine-tune our pre-trained model for speaker adaptation.

\subsection{Base multi-speaker TTS model}\label{speakerrepresentation}
\myparagraph{FastPitch} 
We use FastPitch \cite{lancucki2021fastpitch} as base TTS model. FastPitch model is composed of four components including two feed-forward transformer (FFT) stacks as phoneme encoder and mel decoder, and two convolutional modules as pitch and duration predictor. The encoder operates on the input phoneme tokens $x$ and produces a hidden state $h$ which is used to predict the average pitch of each token $\hat{p}$ and duration $\hat{d}$ by the pitch and duration predictor respectively. The decoder takes the length-regulated hidden representations from the sum of encoder outputs $h$ and pitch $\hat{p}$ to produce the mel-spectrogram sequence $\hat{y}$. To train the pitch predictor, we use the ground-truth pitch $p$, derived using PYIN \cite{mauch2014pyin} and averaged over the input tokens. For duration predictor, we use a learnable aligner from \cite{badlani2022one}. The training loss is composed from MSE between predicted and ground-truth modalities plus the alignment loss:
\begin{equation*}
    L = || \hat{y} - y || ^{2}_{2} + \alpha || \hat{p} - p || ^{2}_{2} + \beta || \hat{d} - d || ^{2}_{2} + \gamma L_{align}.
\end{equation*}

\label{speakerfigure}
\myparagraph{Speaker Representation}The naive way to represent speaker is adding a  speaker embedding table. The limitation of this approach is that it cannot generalize to speakers unseen during training. Therefore, we combine speaker embeddings $(SE_1)$ from look-up table with speaker embeddings $(SE_2)$ obtained from a reference spectrogram and global style tokens \cite{wang2018style} for a particular speaker. The spectrogram is fed to a convolutional RNN reference encoder followed by a multi-head attention layer. The attention module outputs weights to sum the style tokens as a speaker representation embedding. Final speaker embedding $SE_{final}$ is obtained by adding $SE_1$ and $SE_2$. The advantage of this approach is that we can learn tokens without any explicit style or prosody labels but still learn a large range of acoustic expressiveness such as speed, speaker identity, and speaking style.

\myparagraph{Multi-Speaker FastPitch}
We condition FastPitch using speaker representation as an additional input to each FastPitch component: encoder, decoder, pitch predictor, duration predictor, and aligner. Inputs of each components is concatenate with the speaker representation. Following \cite{chen2021adaspeech}, we also leverage conditional layer normalization (CLN) to control our model with the corresponding speaker characteristics. The conditional network consists of two linear layers to project the extracted speaker representation to the scale and bias vector in CLN. We substitute CLN instead of all original LayerNorm layers used in encoder,  pitch and duration predictor, and decoder.

\subsection{Adapter Modules}
\myparagraph{Vanilla Adapter}
Adapters \cite{houlsby2019adapter} are small modules injected between layers of a frozen pre-trained network. During training, the gradient only updates the adapters while other parameters are fixed. The adapter layer generally uses a down-projection feed-forward network $\boldsymbol{(FF_{down})}$ to project the input to a lower-dimensional bottleneck, followed by a non-linear activation function and a up-projection feed-forward network $\boldsymbol{(FF_{up})}$. To stabilize training, a near-identity initialization is required, so the adapter has a skip-connection internally. With the skip-connection, original network can stay unaffected when training starts. In our design, we optionally add dropout and layer normalization as well as the zero initialization of final layer to serve this module as identity operation. Moreover, different from placing adapters inside transformer layers following \cite{houlsby2019adapter}, we propose to insert them after the outputs of each transformer layer. Specifically, we generalize adapters to be inserted after any module.


\myparagraph{Unified Adapters}
Recent work \cite{he2021unified} has proposed a unified framework to integrate previous parameter-efficient modules as adapter variants, such as \textbf{LoRA} \cite{hu2021lora} and \textbf{Prefix Tuning} \cite{li2021prefix}. LoRA injects trainable low-rank matrices into the self-attention network in each transformer layer to update the query and key. The architecture is similar to adapter but without an activation function and with a fixed scaling scalar. Prefix Tuning prepends trainable vectors to the keys and values of self-attention network in each transformer layer. In other words, we concatenate the original key and value matrices with additional prefix vectors and perform multi-head attention as usual. Compared to Adapter, LoRA and Prefix Tuning are only applied to self-attention network in transformers. We also use other simple tuning approach \textbf{BitFit} \cite{zaken2021bitfit}. This approach only updates bias vectors while fixing other parameters in the pre-trained model.

\subsection{Parameter-efficient fine-tuning}
\label{sec:fine-tune}
To fine-tune our frozen pre-trained FastPitch on the new speaker adaptation data, we update only the parameter-efficient modules or speaker-related modules. First, we insert parameter-efficient modules in our pre-trained model. We added vanilla adapter to the phoneme encoder, mel decoder, pitch and duration predictor as well as the aligner. We experimented by applying LoRA and Prefix Tuning to self-attention network in encoder and decoder. BitFit is used in any layers with bias terms. Second, for speaker identity representation, we obtain speaker embedding $(SE_2)$ from reference spectrogram and GST as described in Speaker Representation section of \ref{speakerrepresentation}. We add this speaker embedding with speaker embedding $(SE_1)$ obtained by weighted mean of all pre-trained speaker embeddings from lookup table to form the final speaker embedding $SE_{final}$  as shown in \autoref{fig:fastpitch}. The weights were learnt from gradients during fine-tuning. Third, we also unfreeze the linear layers of scale and bias in each CLN because this module's effectiveness to control speaker identity has been verified in \cite{chen2021adaspeech}.  With a small number of trainable parameters, we can optimize our TTS model for speaker adaptation in a parameter-efficient way.


\section{Experiments and Results}
\subsection{Dataset}
We used  LibriTTS \cite{libritts} for pre-training. We select the top 100 longest-duration speakers with a total of 42.5 hours from the original train-clean-360 set. For the evaluation of speaker adaptation, we create our test set with 10 unseen speakers (5 men and 5 women) from the top longest-duration speakers in the original test-clean set. To simulate few-shot scenario, the test set is composed of 15 minutes data for each speaker. To validate the generalization abilities to multiple acoustic conditions, we experiment on VCTK \cite{vctk} and HiFi-TTS \cite{hifitts} datasets. For each test speaker, we randomly choose 20 unseen utterances to evaluate the adaptation voice quality. 

After the data collection, we normalize and tokenize the raw text sequence into phoneme tokens. Also, we pre-process the speech waveform into mel-spectogram under the sampling rate 22kHz and pre-compute the pitch \cite{mauch2014pyin} and alignment prior \cite{badlani2022one} before training.

\subsection{Experiment Setup}
We pre-train multi-speaker FastPitch for 500 epochs on 8  V100 GPUs with batch size  16 and learning rate $1 \times 10^{-3}$.  

In the fine-tuning stage, we freeze all model parameters and only update the proposed speaker-related and parameter-efficient modules. We train the model as well as our baselines for $\sim$1500 steps with batch size 8, learning rate $2 \times 10^{-4}$ and Adam optimizer on 1 NVIDIA A5000 GPU. The adaptation process may take 10 to 15 minutes depending on the data size. We use HiFi-GAN \cite{kong2020hifi}, trained on mel-spectrograms from pre-trained FastPitch, as the vocoder to convert mel-spectrograms to waveforms. The vocoder was not fine-tuned on the new adapted speakers.

\subsection{Evaluation Metrics}
To measure the voice quality, we conduct both objective and subjective evaluation on the synthesized and ground-truth speech. For objective evaluation, we first calculate the average Speaker Embedding Cosine Similarity (SECS) between the reference and measured audios by a speaker verification model \cite{koluguri2022titanet} to estimate speaker similarity. Further, we compute Conditional Fr\'{e}chet Speech Distance (CFSD) \cite{voicefilter} between the generated speech and actual recording to measure signal quality. Besides, we also evaluate mean square error for pitch (MSE$_P$) and duration (MSE$_D$) to access prosody similarity. The error is computed against ground-truth speech.

For subjective evaluation, we conduct human evaluations with 5-scale MOS (mean opinion score) for naturalness and SMOS (similarity MOS) for speaker similarity on Amazon Mechanical Turk. Each audio sample is rated by 5 workers. We average the scores of all speakers as the final scores. 


\subsection{Results}
We use four voices (two males and two females) to study how different type of adapters and  speaker-related modules described in section~\ref{sec:fine-tune} affects the quality of TTS adaptation. The results are shown in Table~\ref{tab:ablation}. 
\begin{table}[t]
\centering
\resizebox{0.48\textwidth}{!}{
\begin{tabular}{@{}r|cccc|c@{}}
\toprule

                                    \textbf{Method} & \textbf{SECS $\uparrow$}                     & \textbf{CFSD $\downarrow$}                   & \textbf{MSE$_P$ $\downarrow$}     & \textbf{MSE$_D$ $\downarrow$}         & \textbf{Params}          \\ \midrule
\rowcolor{Gray} \multicolumn{6}{c}{\textit{Different parameter-efficient methods}}                                                                      \\
BitFit                              & 0.452             & 56.4          & 71.0  & \textbf{19.1}           & 2.2M            \\ 
PrefixTuning (FFTs)                         & 0.067             & 83.2          & 75.6 & 22.7                  & 0.6M            \\
LoRA (FFTs)                                 & 0.141             & 62.7          & 77.7 & 22.3                    & 2.8M            \\
Adapter (FFTs)                             & \textbf{0.568}   & \textbf{30.0}   & \textbf{65.9} & 21.3           & 2.4M            \\ \midrule

\rowcolor{Gray} \multicolumn{6}{c}{\textit{Different speaker-related modules}}                                                                     \\
Adapter (FFTs/Predictors/Aligner)                     & 0.575          & 28.3          & \textbf{62.7} & 16.9           & 3.5M            \\
+ speaker embedding         & \textbf{0.586}          & \textbf{27.2} & 63.6 & \textbf{16.2}            & 3.5M            \\
+ speaker embedding + CLN   & 0.540        & 46.9          & 66.8 & 17.2                   & 7.8M            \\ 
\midrule
speaker embedding + CLN \cite{chen2021adaspeech}             & 0.513          & 53.5          & 68.0 & 21.7                       & 4.3M            \\
Full fine-tuning                    & 0.604 & 31.0          & 73.8  & 19.7                    & 53.4M           \\

\bottomrule
\end{tabular}}
\vspace{-5pt}
\caption{Comparison of parameter-efficient methods and ablation study of speaker-related modules on objective metrics. We weighted mean looked-up speaker embeddings as new speaker embedding and unfreeze CLN as trainable modules.}
\label{tab:ablation}
\vspace{-10pt}
\end{table}

When inserting parameter-efficient modules in FFTs in encoder and decoder blocks, adapters significantly outperform other approaches on all metrics except duration error. Next we insert adapters into pitch and duration predictors. Adding speaker embedding to adapters inputs improves the performance while unfreezing CLN may degrade the speech metrics. Moreover, adapter get better scores than just using speaker embedding and CLN \cite{chen2021adaspeech}, and they obtain comparable quality to full fine-tuning when using only 7\% parameters.

After validating the best  design for FastPitch adaptation, we study how  much training data is required  for this setting. The  proposed method outperforms full-model fine-tuning for limited data settings. We find that listeners can hardly recognize quality difference even if model was fine-tuned  with only 5 minutes of the speech data for new speaker, although objective metrics still demonstrate the improvements for larger sets, see Table~\ref{tab:duration}. 

Finally, we adapted model on speakers from VCTK and HiFi-TTS dataset to check how the proposed  method performs when speech for new speakers is recorded with different  conditions comparing to the conditions used for LibriTTS dataset used for pre-training. In Table~\ref{tab:dataset},  adapters outperform full fine-tuning on naturalness (MOS) and on speaker similarity (SMOS) we obtain similar performance across different datasets. These results show our framework can be generalized to diverse recording conditions.

\begin{table}[t]
\centering
\resizebox{0.48\textwidth}{!}{
\begin{tabular}{@{}l|c|cc|cccc@{}}
\toprule
                                  \textbf{Method} & \textbf{Trainset, min}  & \textbf{MOS $\uparrow$}  & \textbf{SMOS $\uparrow$}  & \textbf{SECS $\uparrow$}           & \textbf{CFSD $\downarrow$}           & \textbf{MSE$_P$ $\downarrow$} & \textbf{MSE$_D$ $\downarrow$}     \\
\midrule
\multirow{3}{*}{Adapter}
  & 1  &  3.79 ± 0.08   &  3.29 ± 0.10     & 0.421   & 31.4 & 129.5  & 25.2  
  \\
  & 5  &  \textbf{3.90} ± 0.07   &  3.46   ± 0.10   & 0.466   & 26.9 & 108.3  &   21.5  
  \\
  & 15 &  3.85 ± 0.08   &  \textbf{3.48}   ± 0.10   & 0.492   & 24.2 &  \textbf{90.2}   & 19.1  
  \\
  & 60 &  3.90 ± 0.07  &  3.35 ± 0.10   & \textbf{0.520}   & \textbf{23.2} & 119.6  & \textbf{18.3}  
  \\
\midrule
\multirow{3}{*}{Full fine-tuning} 
  & 1  &  3.44 ± 0.09   &  3.25 ± 0.10     & 0.461 & 35.4 & 135.6 & 30.4 
  \\
  & 5 &  3.68 ± 0.08   &  3.42 ± 0.10    & 0.522 & 26.2  & 102.6 & 25.8         \\
  & 15 &  \textbf{3.72} ± 0.08   &  3.39 ± 0.10    & 0.537 & 25.4          & \textbf{90.5} & 24.6 
  \\
  & 60 &  3.71 ± 0.08   &  \textbf{3.46} ± 0.10    & \textbf{0.542} & \textbf{22.1} & 106.4 & \textbf{22.9}  \\
\bottomrule
\end{tabular}}
\vspace{-5pt}
\caption{Comparison of different amount of training data on both subjective and objective metrics. We fine-tune adapters and the weights to sum looked-up speaker embeddings in all FastPitch components as the adapter results shown here. Note that we omit $\times 10^{-3}$ in reported MSE scores for simplicity.
}
\label{tab:duration}
\vspace{0pt}
\end{table}
\begin{table}[t]
\centering
\resizebox{0.48\textwidth}{!}{
\begin{tabular}{@{}lcccccc@{}}
& \multicolumn{3}{c}{\textbf{MOS $\uparrow$}}                      & \multicolumn{3}{c}{\textbf{SMOS $\uparrow$}}   \\
\toprule 
                                                     \textbf{Method} & \multicolumn{1}{|c}{\textbf{LibriTTS}} & \textbf{VCTK} & \textbf{HiFi-TTS} & \multicolumn{1}{|c}{\textbf{LibriTTS}} & \textbf{VCTK} & \textbf{HiFi-TTS}  \\
\midrule
Recording                                                  &  \multicolumn{1}{|c}{4.16 ± 0.05}          &       4.08 ± 0.05     &     4.12 ± 0.05   &  \multicolumn{1}{|c}{3.90 ± 0.07}          &     3.80 ± 0.07       &     3.71 ± 0.08 \\
\midrule
Adapter                                              &  \multicolumn{1}{|c}{\textbf{4.10 ± 0.05}}          &      \textbf{3.89 ± 0.06}      &   \textbf{3.87 ± 0.06}     &  \multicolumn{1}{|c}{\textbf{3.44 ± 0.08}}          &     3.28 ± 0.08       &    3.27 ± 0.08  \\
Full fine-tuning                                     &  \multicolumn{1}{|c}{3.96 ± 0.06}          &      3.88 ± 0.06      &     3.75 ± 0.07   &  \multicolumn{1}{|c}{3.40 ± 0.08}          &     \textbf{3.33 ± 0.06}       &   \textbf{3.34 ± 0.08}   \\
\bottomrule
\end{tabular}}
\caption{Comparison of datasets with different acoustic conditions on subjective metrics.}
\label{tab:dataset}
\vspace{-5pt}
\end{table}



\section{Conclusion}
In this work, we propose parameter-efficient method for adaptation of multi-speaker TTS models to new speakers. The new speaker adaptation is based on adding small adapter modules to base model. We keep weights of base model frozen, and only parameters of adapters are fine-tuned on new speaker data. 
The experiments show proposed adaptation method achieves  similar speech naturalness, speaker and prosody similarity while requires significantly less compute. It also performs well even in low-data regime.  

The model is open-sourced in Nemo toolkit.\footnote{https://github.com/NVIDIA/NeMo}

\bibliographystyle{IEEEbib}
\bibliography{refs}

\end{document}